# Use of Ghost Cytometry to Differentiate Cells with Similar Gross Morphologic Characteristics

**Authors:** Hiroaki Adachi[1], Yoko Kawamura[1], Keiji Nakagawa[1], Ryoichi Horisaki[3,4], Issei Sato[1,2,5], Satoko Yamaguchi[2], Katsuhito Fujiu[2], Kayo Waki[2], Hiroyuki Noji[2], Sadao Ota[1,2,*]

**Affiliations:**

[1]Thinkcyte Inc., 7-3-1 Hongo, Bunkyo-ku, Tokyo 113-8654, Japan.

[2]University of Tokyo, 7-3-1 Hongo, Bunkyo-ku, Tokyo 113-8654, Japan

[3]Department of Information and Physical Sciences, Graduate School of Information Science and Technology, Osaka University, 1-5 Yamadaoka, Suita, Osaka 565-0871, Japan.

[4]JST, PRESTO, 4-1-8 Honcho, Kawaguchi-shi, Saitama 332-0012, Japan.

[5]RIKEN AIP, Nihonbashi 1-chome Mitsui Building, 15th floor, 1-4-1 Nihonbashi, Chuo-ku, Tokyo, 103-0027, Japan.

*Correspondence to:  sadaota@solab.rcast.u-tokyo.ac.jp

**Abstract**:

Imaging flow cytometry shows significant potential for increasing our understanding of heterogeneous and complex life systems and is useful for biomedical applications. Ghost cytometry is a recently proposed approach for directly analyzing compressively measured signals, thereby relieving the computational bottleneck observed in high-throughput cytometry based on morphological information. While this image-free approach could distinguish different cell types using the same fluorescence staining method, further strict controls are sometimes required to clearly demonstrate that the classification is based on detailed morphologic analysis. In this study, we show that ghost cytometry can be used to classify cell populations of the same type but with different fluorescence distributions in space, supporting the strength of our image-free approach for morphologic cell analysis.

**Main**:

Ghost cytometry (GC) is a machine learning-integrated method that directly analyzes compressed morphologic information of cells without image production (*1*). In a previous study, we classified cells exhibiting subtle differences in their two-dimensional images (*2*) that were not easily distinguishable by the human eye. The capability of GC to classify such similar cell morphologies raised the question of whether the direct analysis of the waveforms could be based on non-morphological information (*3*). Compressively-measured signals do contain morphologic information from cells. However, it is not straightforward to exclude the possibility that the signal also encodes non-morphological information such as the speed of cells in flow, which may be convolved with morphological information such as cell size. Additionally, image-based cytometry is expected to be capable of analyzing in more detail than basic morphological information such as size and fluorescence intensity.

Here, we show that GC can classify cells based on morphological information in a detailed manner. We used the same cell type exhibiting apparently different image patterns of fluorescence in the same color channel, thereby minimizing the possible non-morphological factors specific to different cell types. Furthermore, in contrast to the high performance of GC-based classification, we confirmed that the cells were not well-separated by analysis based on a pair of representative features which we suppose as *basic* morphological information in this study and obtained by using a conventional flow cytometer or simply by extracting them from the GC waveforms, respectively. The results showed that GC classifies cells based on morphologic information in more detail than the analysis based on the two features.

**Figure 1A** shows a schematic of the optical setup in our GC analyzer, which is equipped with two different continuous wave lasers. A blue laser (488 nm) forms a structured illumination pattern using a diffractive optical element and objective lens. When cells pass through the pattern projected inside a glass flow cell, the excitation pattern of light generates temporally modulated waveforms of green fluorescence to be classified. A red laser (637 nm) is used to obtain standard forward scattering (FSC) signals, which trigger the acquisition of a waveform set. Red fluorescence signals were used to label the waveforms for training of the classifier and validate the GC-based classification results. The green and red fluorescence signals were collected through the objective lens used for excitation (Obj 1), while the FSC signals were collected through another objective lens (Obj 2). The set of these optical signals collected from each cell was simultaneously detected by each photomultiplier tube (PMT), and then recorded by using an analog-to-digital converter (ADC) board with a trigger condition applied on the FSC signals, as shown in **Fig. 1B**. Cells were hydrodynamically focused into a single narrow stream and aligned along the structured illumination. The flow was controlled at a constant rate and the waveform width was maintained at shorter than 100 µs (**Fig. 1B**). This temporal waveform width corresponds to a theoretical maximum throughput of 10,000 events/s based on an assumption of equal intervals between cells.

While this and other similar setups allow for the production of fluorescence images based on the concept of computational ghost imaging (*1, 2, 4-6*), in this study, we perform the direct analysis of the waveforms using a supervised machine learning model based on a support vector machine (SVM). In training the model using the experimental setup, we first prepared a training dataset by simultaneously acquiring pairs of the green fluorescence waveform and the red fluorescence as a label from each cell. Using many sets of these labeled waveforms, we next trained the machine classifier which, in turn, predicted the red fluorescence label from the green fluorescence waveform.

In the experiment, a population of cultured Raji cells (Burkitt's lymphoma cell line) was separated and stained with either Calcein-AM for green fluorescence staining of the cytoplasm or MitoTracker Green FM for green fluorescence staining of the mitochondria. **Figure 1C** shows example images of the two cell populations acquired with a commercial imaging flow cytometer (Amnis® FlowSight®, Luminex, Inc.). The staining conditions were optimized such that the two populations showed similar total intensities in the green fluorescence waveforms (**Fig. 2A**). Additionally, only the cells stained with MitoTracker Green FM were further labeled with CellMask Deep Red Plasma membrane Stain, wherein the red fluorescence was used to label the green fluorescence waveforms and to validate the GC-based classification results. In the GC-based measurements, the differently stained Raji cells were mixed at an approximately equal ratio and immediately introduced into the flow cell.

In **Fig. 1B**, the left and right panels show example sets of the signals obtained for the cell mixture. The signal sets were grouped using a gating condition defined based on the scatter plots showing the peak intensity of the red fluorescence label against the total intensity of the green fluorescence waveform obtained for the cell mixture (rectangles in **Fig. 2A**). This gating condition was used in preparing a data set of the waveforms with validation labels for training and testing of the model. To the waveform set which was lastly smoothed and normalized in a cell-type independent manner, we performed the SVM-based classification of the two populations. 1,000 of the labeled waveforms from each cell population were used for training and 100 of those were used from the rest of the population for testing, respectively.

**Figure 2B** is a confusion matrix obtained by comparing the GC-based classification results and validation results based on quantification of the red labels. **Figure 2C** is a receiver operating characteristic (ROC) curve measured from the mixed populations, recording > 0.999 as the area under the curve (AUC) score. The inset in **Fig. 2C** is a histogram of scores obtained by applying the trained SVM-based classifier to the waveform signals, and its colors were obtained by applying the gating condition shown in **Fig. 2A**. Thus, GC accurately classified the cell populations containing the same type of cells with different stains, which showed apparently different image patterns (**Fig. 1C**). We believe that controlling the cell type minimizes the possible effects of non-morphological factors specific to different cell types on the classification performance of GC, including the effects of velocity differences due to variance in their deformability (*7*).

To study which type of morphological information affects GC, we used a pair of quantities representing basic morphological information to classify the same cell mixture, specifically, the heights of FSC and side scattering (SSC) signals (*8*), and width and height of the waveforms (*1,3*). While we note there may be other quantities of basic morphology, the parameters we chose in this study could be readily extracted from the measured signals. The FSC and SSC signal are known to indicate the size and internal complexity (i.e. granularity) of the cells, respectively, and were a pair of quantities that could be obtained simultaneously for each cell using a commercial flow cytometer (JSAN, Bay Biosciences). We suppose that the waveform width and height indicate the velocity and size of cells, and the total amount of fluorescence molecules within the cells, respectively, and were a pair of quantities that could be obtained simultaneously for each cell using our GC analyzer setup. **Figure 3A** shows the height of FSC signals against that of SSC signals obtained with the commercial flow cytometer for the same cell mixture, which confirmed the highly overlapped distribution of the two populations. We performed the SVM-based classification of the two populations using the FSC and SSC heights, which were normalized in a cell-type independent manner. 1,000 of the pairs of the FSC and SSC heights from each cell population were used for training while a 100 of those were used from the rest of the population for testing, respectively. Finally, we obtained a limited value of 0.670 as the AUC score.

**Figure 3B** shows a plot of the height of the waveforms against the width of the waveforms obtained by using our GC analyzer setup and applying the gating condition shown in **Fig. 2A**, with the overlapped distribution shown for the two populations. When we used only the height and width of the waveforms for training and testing of the SVM model, we obtained 0.7056 as the AUC score. To control for differences in the waveform height or width due to possible variance in cell size, velocity, and fluorescence intensity, we applied the SVM models trained using the 2,000 cells used in **Fig. 2B** and **2C** for analysis of a restricted data set consisting of 657 cells with similar waveform width and height values (included inside the red rectangle in **Fig. 3B**). Importantly, the AUC for the waveform-based analysis of cells with similar waveform width and height was 1.000, while that obtained by two feature-based analysis was only 0.5748 (ROC curves in **Fig. 3C**). Thus, analysis of all information encoded in the compressive waveform improved classification performance compared to analysis based on only the two features representing basic morphological information, supporting that GC classifies cells based on morphological information in a detailed manner.

Additionally, for the purpose of comparison and confirming that the cell populations were morphologically separable, we used a library of two-dimensional (2D) fluorescence images of the two cell populations (images in **Fig. 1C** are from this library), obtained with a commercial imaging flow cytometer (20x objective lens, FlowSight), for training and testing the SVM model. We preprocessed the data set in this image library using IDEAS® software (Luminex, Inc.) to identify a population of focused single cells (*9*) before cropping raw images around the center of mass into images of 30 × 30 pixels, and randomly rotating them with a multiple of 90 degrees, wherein error images with a saturated pixel or cut-off images of cells were removed. To the training and test image datasets of 1,000 and 100 single cells for each cell population, respectively, which was lastly normalized in a cell-type independent manner, we performed an image-based SVM classification. **Figure 4A** shows a confusion matrix obtained from the SVM-based 2D image analysis. **Figure 4B** shows the ROC curve obtained from this analysis and its AUC score was 0.9793. The inset in **Fig. 4B** is a histogram of scores obtained by applying the SVM-based classifier to the 2D images while the colors were assigned to each staining method. Slight difference in the AUCs between the GC- and image-based analysis could arise from various possibilities including differences in the fluidic and/or optical methods adopted for measuring the morphological information of cells. These results show that the classification performance based on the 2D images was comparable to that based on the GC waveforms.

In conclusion, we show that direct analysis of the compressive waveforms in GC can accurately distinguish the cell populations exhibiting differences in the spatial distributions of fluorescence within the cell. The discrimination performance of this analysis was superior to one employing the FSC and SSC, or another employing the height and width of the waveforms, respectively. This experimental result is consistent with the mathematical proof that classification using the GC methodology is based on the image information of cells (*10*).

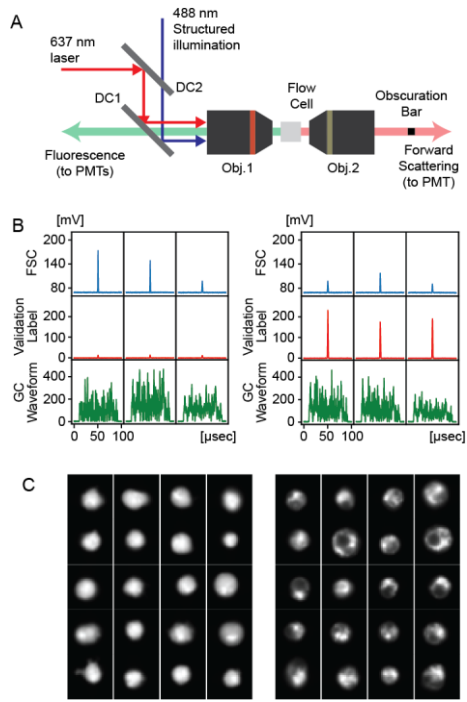

**Figure 1**
**Ghost Cytometry-based measurement of a cell type exhibiting different image patterns.**
(**A**) In the ghost cytometry (GC) analyzer, cells were hydrodynamically focused and illuminated by two lasers to record three types of signals from each cell: one was a structured illumination to acquire temporally modulated fluorescence waveforms (green channel), while another was elliptically focused illumination to acquire forward scattering signals and fluorescence signals (red channel) used to label the waveforms and validate the waveform-based classification results. (**B**) Example sets of optical signals detected simultaneously for each cell using three photomultiplier tubes in parallel. Left panel shows three sets of signals obtained from Raji cells stained with Calcein-AM only (bottom row), showing weak signals in the red channel (middle row). Right shows three sets of signals obtained from Raji cells stained with MitoTracker Green FM and CellMask Deep Red Plasma membrane Stain, showing strong signals in the red channel. From these data sets, quantification of the fluorescence intensity in the red channel gives labels with each associated waveform. This training data set of the labeled waveforms enables training of a machine learning-based model which, in turn, predicts the label from the waveform. (**C**) Left shows example fluorescence images of Raji cells with their cytoplasm stained with Calcein-AM and right is those with their mitochondria stained with MitoTracker Green FM, acquired with a commercial imaging flow cytometer (20x objective lens, FlowSight).

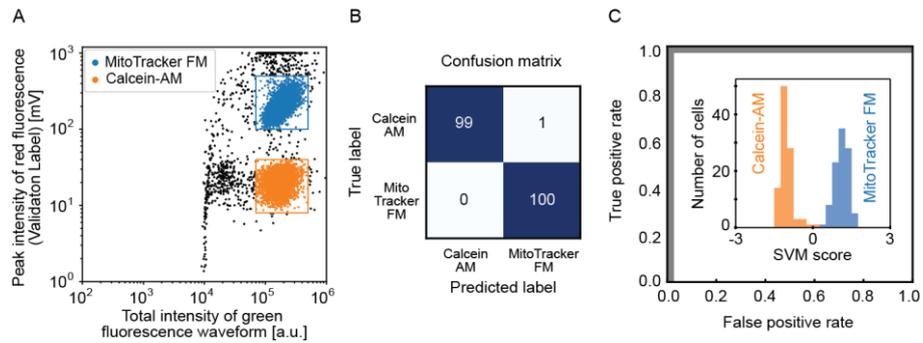

**Figure 2**
**Cell classification by direct analysis of GC waveforms.**
(**A**) Scatter plot of the peak intensity of validation labels of red fluorescence against the total intensity of the green fluorescence waveforms obtained for the two populations. Rectangles are gating conditions used to select and label the waveform data for training the cell classifier and to validate the GC-based classification results. Orange and blue plots were selected as positively and negatively labeled cells from original black plots of a whole cell population recorded. (**B**) Confusion matrix obtained by comparing the classification result using GC and that using validation labels. (**C**) ROC curve obtained for the GC-based classification result compared to the validation labels, showing an AUC score > 0.999. The inset in **C** is a histogram of scores obtained when the trained SVM-based classifier was applied to the waveforms with colors obtained by applying the gating condition shown in **A**.

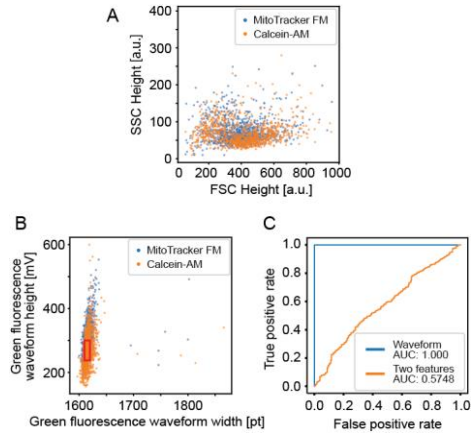

**Figure 3**
**Cell classification by analysis of two features representing *basic* morphological information.**
(**A**) Scatter plot of the forward scattering and side scattering signals obtained by using a conventional flow cytometer (JSAN). (**B**) Scatter plot of height and width distribution of the waveforms obtained by using the GC analyzer setup shown in **Fig. 1A** and applying the gate condition shown in **Fig. 2A**. The rectangle shows an area where the two populations overlapped well for cells classified in **C**. (**C**) ROC curves obtained by using the training datasets of the same 2,000 cells used in **Fig. 2B** and **2C**, but with classifications performed for 657 cells inside the rectangle region of **Fig. 3B**. Blue and orange curves are ROC curves measured by analysis of the whole waveform and that based on the height and width of the waveform, respectively.

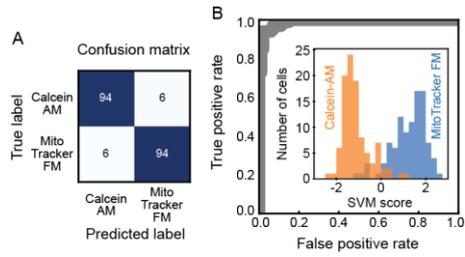

**Figure 4**
**Cell classification using two-dimensional (2D) fluorescence images.**
(**A**) Confusion matrix obtained by a SVM-based cell classification based on the 2D fluorescence image library, obtained by flowing the two populations separately into the commercial imaging flow cytometer (FlowSight). (**B**) ROC curve obtained for the two cell populations, recording 0.9793 of an AUC score. The inset in **B** is a histogram of scores obtained by applying the classifier to the 2D images while its colors were assigned to each staining method.

## Methods
### Hardware setups (optics, electronics, and fluidics)
In the optical system, a blue laser (Voltran/Stradus, USA) and a red laser (Voltran/Stradus, USA) were combined using a dichroic mirror DC2 (NFD01-633, Semrock) in Fig. 1A and then illuminated cells flowing in a glass flow cell (Hamamatsu Photonics K.K.) through an objective lens (Obj 1, UPLSAPO 20x, Olympus). Emitted fluorescence photons passed through Obj 1 and DC1 (Di03-R405/488/561/635-t3, Semrock, in Fig. 1A), and were separated by other dichroic mirror DC3 (Di02-R561, Semrock, not shown in Fig. 1A) and DC4 (Di02-R635, Semrock, not shown in Fig. 1A) to be detected by two PMTs (Hamamatsu Photonics K.K.) through each bandpass filter (FF03-525/50-25, Semrock, and FF01-680/42-25, Semrock, not shown in Fig. 1A), separately. Forward scattering photons were collected by another objective lens (Obj 2, UPLFLN 10x2, Olympus), spatially filtered with an obscuration bar, and detected by a PMT through a bandpass filter (FF01-640/40-25, Semrock, not shown in Fig. 1A). A PMT of 10 MHz with built-in amplifier was used for detecting green fluorescence signals while PMTs of 1 MHz with built-in amplifiers were used to detect FSC and red fluorescence signals. These signals detected by PMTs were recorded by using an ADC (M2i.4932, Spectrum) board based on a trigger condition applied to the FSC signals (Fig. 1B). In the fluidic system, a syringe pump (Legato 111, KD Scientific) was used to introduce a sample fluid while a pressure pump was used to introduce a sheath fluid. Cells were hydrodynamically focused into a single narrow stream in the glass flow cell. The rates of the sample and sheath fluids were maintained constant at 20 µL/min and 15 mL/min, respectively.

### Cell Staining
Raji cells (Burkitt's lymphoma cell line) were provided by JCRB Cell Bank and cultured in RPMI1640 medium supplemented with 20% Fetal Bovine Serum. All the staining processes were performed at room temperature. Raji cells in the medium were pelleted by centrifugation and resuspended in phosphate buffered salt (PBS). Two samples were prepared with each containing about $5 \times 10^6$ cells in suspension. The cells were pelleted by centrifugation and incubated with MitoTracker Green FM (diluted 1:50 in PBS from 1 mM MitoTracker Green FM in dimethyl sulfoxide (DMSO)) and with Calcein-AM (diluted 1:500 in PBS from Calcein-AM (1mg/mL) in DMSO), respectively. After 30 min incubation, the cells were washed in PBS and pelleted by centrifugation. Only the cells stained by MitoTracker Green FM were additionally labeled in red fluorescence by 15 min incubation with CellMask Deep Red Plasma membrane Stain (diluted 1:500 in PBS from CellMask Deep Red Plasma membrane Stain (5mg/mL) in DMSO). On the other hand, the cells stained by Calcein-AM were incubated for 15 min in PBS containing 0.2% DMSO so that the total time of exposure to DMSO was equal between the two samples. After staining, the cells were washed in PBS 3 times by centrifugation. The cell concentration was finally adjusted to around $2 \times 10^6$ cells/mL in PBS for FlowSight analysis and $2 \times 10^5$ cells/mL in PBS for GC analysis, respectively.

### Analysis
All SVM used kernel method algorithm. All hyperparameter were selected by 10-fold cross-validation of the ROC-AUC. All Figures 3 show the scatters plots of example 1,000 data points for each cell population.

In Fig. 2, from the sets of optical signals measured for each event using three PMTs in the GC analyzer setup, we plotted scatters of peak red fluorescence intensities against total intensities of the green fluorescent waveforms. The gating conditions shown as two rectangles in Fig. 2A were then defined to select positively labeled and negatively labeled waveforms, respectively. After preparing a library of the waveforms, which were 2,048 points in a temporal domain, based on the gating condition, we divided this original library into a train-library and test-library using a stratified splitting. For training the SVM model, 1,000 positive waveforms and 1,000 negative waveforms were randomly selected from the train-library as a training data set. For testing the model, 100 positive and 100 negative waveforms were randomly selected from the test-library as a test data set. After smoothing and normalization, we trained and tested the SVM model (Fig. 2B and 2C).

In Fig. 3A, we obtained height of FSCs and SSCs for each cell by flowing the two populations into a conventional flow cytometer (JSAN), separately (so we have all labels of cells). The measured data was randomly separated using stratified splitting into a training-library and test-library. 1,000 positive cells and 1,000 negative cells were used from the training-library as a training data set. For testing the model, 100 positive cells and 100 negative cells were used from the test-library as a test data set. After normalization, we trained and tested the SVM model.

In plotting Fig. 3B, we applied a threshold value of 17.0822 mV to the waveforms to obtain the waveform width and measured the maximum value of the waveform to obtain the waveform height. In performing the SVM analysis for Fig. 3C, While the same 2,000 cells used as a training data set in Fig. 2B and 2C were used for training, a restricted data set consisting of 657 cells (the cell population included both in a red rectangle in Fig. 3B as well as in the test-library) was used for testing. After normalization, we trained and tested the SVM model.

In Fig. 4, After selecting focused single cells from the raw images obtained with a commercial image flow cytometer (FlowSight), we prepared an image library by cropping the cell images into $30 \times 30$ pixels, removing error images with a saturated pixel or cut-off images of cells, and randomly rotating them with a multiple of 90 degrees. For training the SVM model, 1,000 images wherein the cells were stained with Calcein-AM and 1,000 images wherein the cells were stained with MitoTracker Green FM were randomly selected as a training data set. For testing the model, 100 images for each stain were randomly selected as a test data set. After normalization, we trained and tested the SVM model.